\documentclass[preprint,showpacs,preprintnumbers,amsmath,amssymb]{revtex4}

\usepackage{graphicx}
\usepackage{dcolumn}
\usepackage{bm}

\begin{document}


\title{Single and pair production of heavy leptons in $E_6$ model}

\author{A. T. Alan}
\email{alan_a@ibu.edu.tr}
\author{A. T. Tasci}
\email{tasci_a@ibu.edu.tr}
\author{N. Karagoz}
 \email{karagoz_n@ibu.edu.tr}
\affiliation{Abant Izzet Baysal University, Department of Physics,
14280 Bolu, Turkey}

\pacs{12.60.-i, 13.66.De, 14.60.-z}

\date{\today}

\begin{abstract}
We investigate the single and pair production of new heavy leptons
via string inspired $E_{6}$ model at future linear colliders.
Signal and corresponding backgrounds for these leptons are
studied. We have found that single production of heavy leptons is
more relevant than that of pair production, as expected. For a
maximal mixing value of 0.1, the upper mass limits of 2750 GeV in
the single case and 1250 GeV in the pair production case are
obtained at $\sqrt{s}=3$ TeV collider option.
\end{abstract}
\maketitle

\section{introduction}

The Standard Model (SM) describes many phenomena up to the
energies that can be reached today. However, some problems like
mass hierarchy and the number of fermion generations can not be
addressed by the SM. Many models, which include new fermions and
interactions, have been developed to deal with the mentioned
shortcomings of the SM. The string inspired $E_6$ model is a well
motivated one which includes extra gauge bosons and new fermions
assigned to the 27-dimensional representation
\cite{Hewett:1988xc}. In the search for extensions of the SM the
new heavy leptons play an important role. Many analysis have been
done for the production of heavy leptons at future $e^-e^+$
\cite{Almeida:1990ay, Almeida:1994ad, Almeida:2000yx,
Almeida:2003cy}, at hadron \cite{Frampton:1992ik, Coutinho:1998bu,
CiezaMontalvo:2002gk} and also at $ep$ colliders
\cite{Rizzo:1986wf, Alan:2004rv}. The experimental upper bounds
for the heavy lepton masses were found to be 44 GeV by OPAL
\cite{Akrawy:1990dx}, 46 GeV by ALEPH \cite{Decamp:1991uy} and 90
GeV by H1 \cite{Ahmed:1994yi} Collaborations. This means that,
heavy leptons could be detected at future high energy colliders.

We have previously considered the single production of new heavy
leptons via string inspired $E_{6}$ model in $ep$ collisions
\cite{Alan:2004rv}. In this study, we consider both the single and
the pair production of heavy leptons separately by using the same
model and taking into account the signal and background events at
future linear colliders. The main parameters of these collider
options were taken from Refs. \cite{Accomando:2004sz, lc} and
displayed in Table \ref{tab:table1} .

\section{single production of heavy leptons}
The single production of heavy leptons $L$, in $e^-e^+$ collisions
occur through the $s$ and $t$ channel processes $e^-e^+\rightarrow
Le^+$ caused by the flavor changing neutral current (FCNC)
Lagrangian:
\begin{eqnarray}\label{e1}
\mathcal{L}_{\mathtt{nc}}=g_{z}\sin\theta_{mix}\psi_{L}\gamma^{\mu}(1+\gamma_{5})\psi_{e}Z^{\mu}+h.c.\end{eqnarray}
where $\sin\theta_{mix}$ are the mixing angles between right
handed components of the ordinary and new heavy charged leptons.
The order of the mixings of the ordinary and heavy leptons are
known to be $\sin^2\theta_{mix}\approx10^{-2}-10^{-3}$, which
comes from low energy phenomenological calculations and the high
precision measurements of the $Z$ properties at linear colliders
\cite{Bamert:1995wy, Almeida:2000pz, Nardi:1994iv}. We use the
parameter $b_{lLZ}$ to denote the mixing angles and take $0.1$ as
an upper value in the numerical calculations.

The differential production cross section takes the form,
\begin{eqnarray}\label{e2}
    \frac{d\sigma}{dt}&=&\frac{\pi\alpha^2b_{lLZ}^2}{\sin^4\theta_w\cos^4\theta_w s^2}\left[\frac{4(m^2-s-t)(s+t)(s-M_Z^2)(t-M_Z^2)(a_e-v_e)^2}
    {\left[(s-M_Z^2)^2+M_Z^2\Gamma_Z^2\right]\left[(t-M_Z^2)^2+M_Z^2\Gamma_Z^2\right]}\right.\nonumber\\
&&\left. -\frac{(m^2-s-t)(s+t)(a_e-v_e)^2+t(m^2-t)(a_e+v_e)^2}
    {(s-M_Z^2)^2+M_Z^2\Gamma_Z^2}\right.\nonumber\\
&&\left.+
\frac{\left[\left[(s+t)^2+s^2-(2s+t)m^2\right](a_e^2+v_e^2)+2t(m^2-2s-t)a_ev_e\right]}{(t-M_Z^2)^2+M_Z^2\Gamma_Z^2}\right]
\end{eqnarray}
where $a_e=-\frac{1}{2}$ and $v_e=-\frac{1}{2}+2\sin^2\theta_w$,
$\Gamma_Z$ is the decay width and $M_Z$ is the mass of $Z$ boson,
$s$ and $t$ are Mandelstam variables.

\begin{figure}
\includegraphics[width=7cm,
angle=-90]{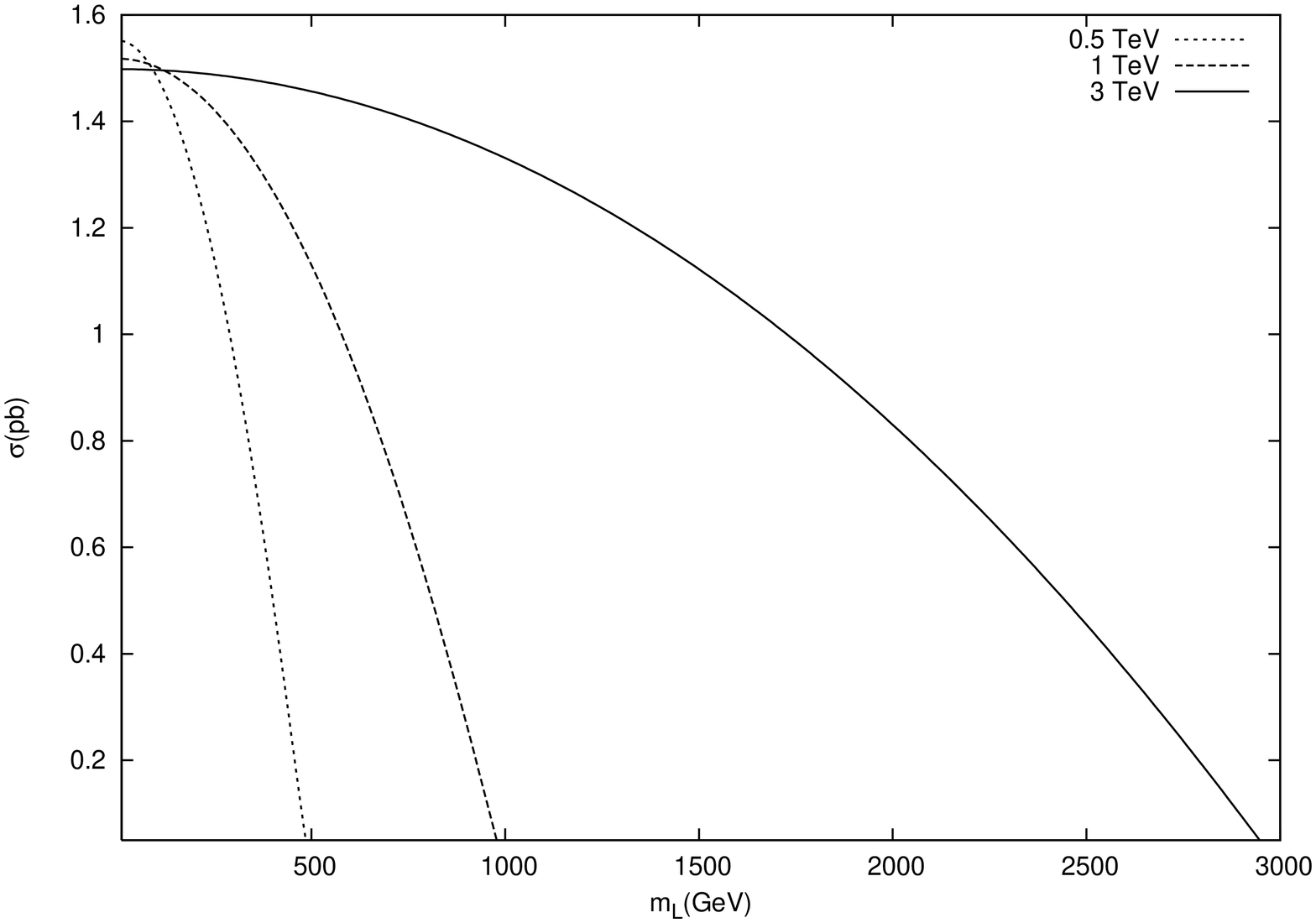} \caption{\label{fig:f1}The total cross
sections as function of the heavy lepton masses, for the single
production of heavy lepton with $\sqrt{s}=$0.5 TeV, 1 TeV and 3
TeV.}
\end{figure}

In Fig.~\ref{fig:f1}, we display the total cross sections as
function of the heavy lepton masses, for the three center of mass
energies of the proposed options. After their production, the
heavy leptons will decay via the neutral current process
$L\rightarrow lZ$, where $l=e, \mu, \tau$. The branching ratios
for these processes would be around $33\%$ for each channel.

In Tables~\ref{tab:t2}, \ref{tab:t3} and \ref{tab:t4}, we
presented the single production cross sections ($\sigma\times
\mathrm{BR}_1$), signal and background cross sections depending on
the heavy lepton mass $m_L$, for 0.5, 1 and 3 TeV energy $e^-e^+$
colliders, respectively. The branching ratios $\mathrm{BR}_1$ and
$\mathrm{BR}_2$ refer to $\mathrm{BR}(L\rightarrow Ze)$ and
$\mathrm{BR}(Z\rightarrow e^+e^-,~\mu^+\mu^-)$. The significance
of signal and background is defined as $S/\sqrt{S+B}$, here $S$
and $B$ are the signal and background number of events. The total
decay widths of the heavy leptons are given in the last column of
the tables. As seen from these tables, the $S/\sqrt{S+B}$ values
are higher than five, which is enough for observability, up to the
center of mass energies of the colliders. Single production of
heavy lepton is feasible up to the center of mass energies of the
$e^-e^+$ colliders even with smaller mixing coupling values. For
example, if we take at least 10 signal events and
$S/\sqrt{S+B}\geq 5$ as discovery criteria, the ILC with
$\sqrt{s}=$0.5 TeV can probe mixing values of $b_{lLZ}=0.032$ for
350 GeV leptons. The same couplings can be probed at $\sqrt{s}=$1
and $\sqrt{s}=$3 TeV for even greater masses such as 800 and 2750
GeV.

\begin{figure}
\includegraphics[]{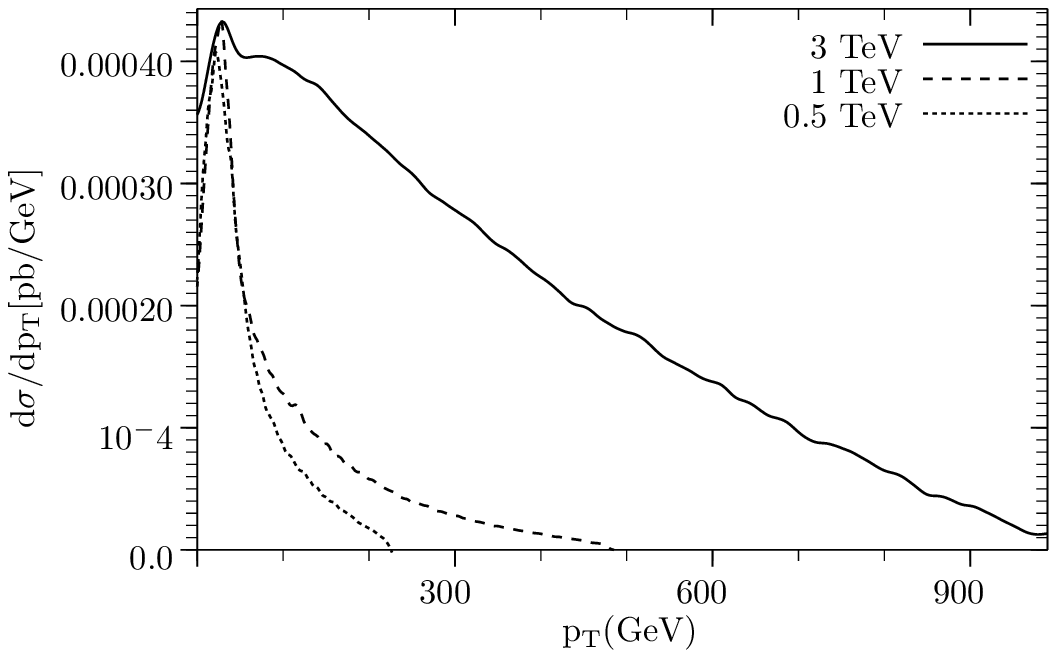}
\caption{\label{fig:f2}$p_T$ distribution of the background at
$\sqrt{s}=$0.5 TeV, 1 TeV and 3 TeV.}
\end{figure}

\begin{figure}
\includegraphics[]{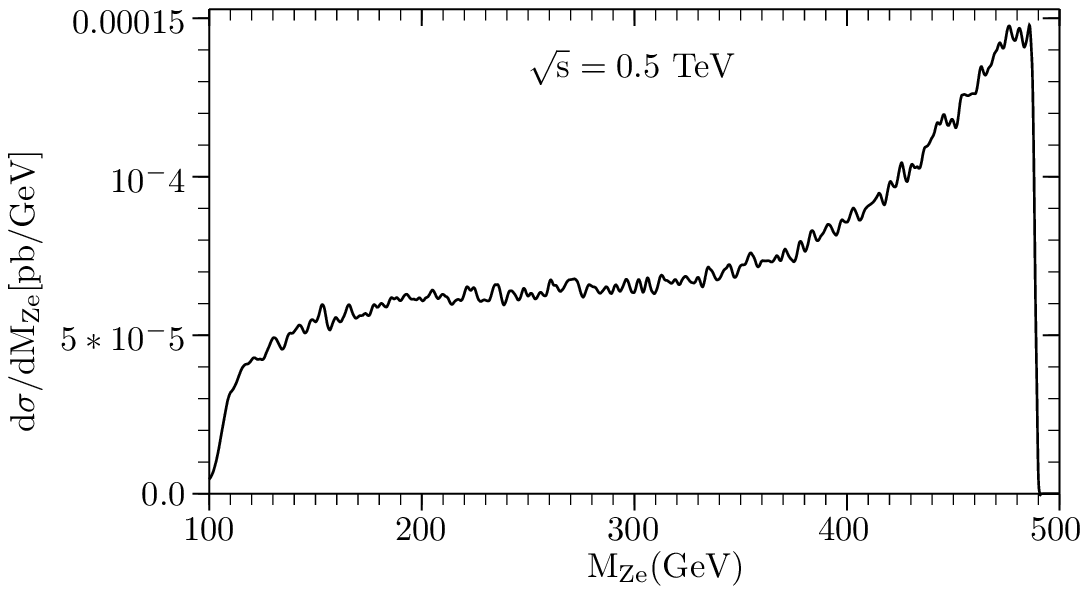}
\caption{\label{fig:f3}The invariant mass distribution of the
$Ze^-$ system for the background at$\sqrt{s}=$0.5 TeV.}
\end{figure}

\begin{figure}
\includegraphics[]{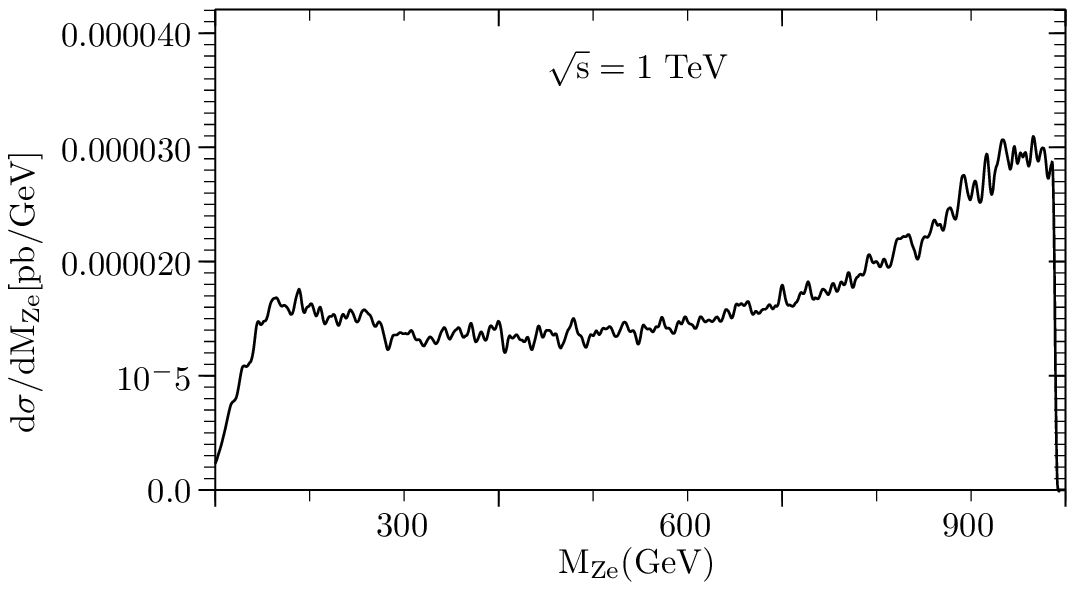}
\caption{\label{fig:f4}The invariant mass distribution of the
$Ze^-$ system for the background at$\sqrt{s}=$1 TeV.}
\end{figure}

\begin{figure}
\includegraphics[]{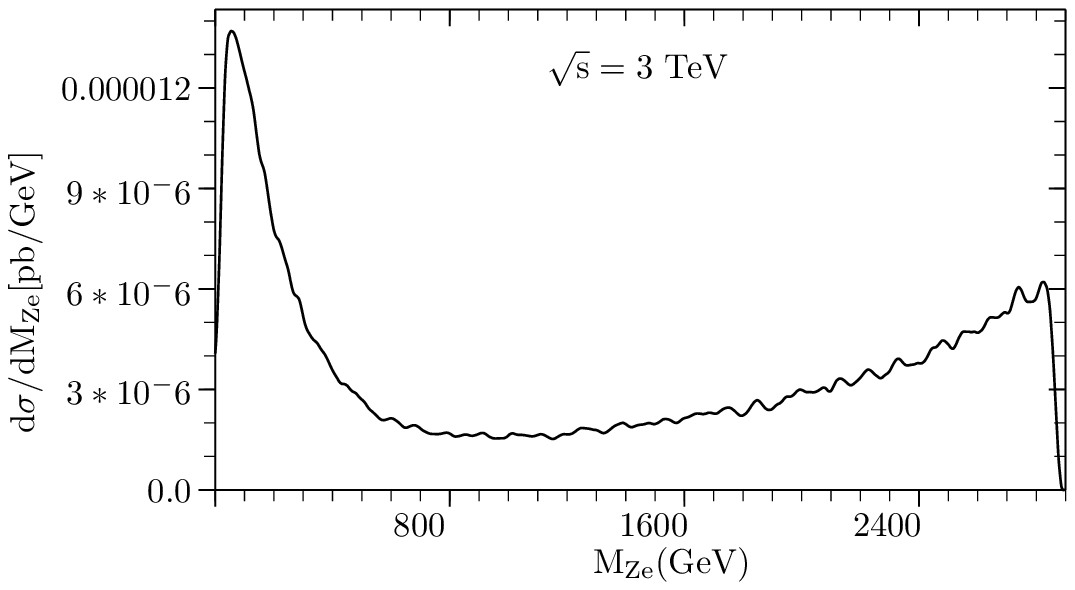}
\caption{\label{fig:f5}The invariant mass distribution of the
$Ze^-$ system for the background at$\sqrt{s}=$3 TeV.}
\end{figure}

We applied a cut of $|m_{Ze}-m_L|<10$ GeV in order to form the
signal and reduce the background for the SM background process
$e^-e^+\rightarrow e^-Ze^+$. Fig.~\ref{fig:f2} shows the $p_T$
distributions at three different linear colliders. In
Figs.~\ref{fig:f3}, \ref{fig:f4} and \ref{fig:f5}, we give the
invariant mass distributions $m_{Ze}$ with cut $p_T^{e^-,j}>10$
GeV at $\sqrt{s}=$0.5, 1 and 3 TeV, respectively.
Figs.~\ref{fig:f3} and \ref{fig:f4} have an increasing character
for a cut 10 GeV interestingly, but lose this character for higher
cuts. For instance, all of the three distributions in
Figs.~\ref{fig:f3}-\ref{fig:f5} have decreasing characters with
Jacobian peaks around 150-200 GeV for a cut value of 80 GeV.

\section{pair production of heavy leptons}

Pair production of heavy leptons in $E_6$ occur through the
$t$-channel flavor changing neutral current process
$e^-e^+\rightarrow L^-L^+$, and the differential cross section for
this process is given by,
\begin{equation}\label{e3}
    \frac{d\sigma}{dt}=\frac{4\pi\alpha^2b_{lLZ}^4}{s^2M_Z^4\left[(t-M_Z^2)^2+M_Z^2\Gamma_Z^2\right]}\left[(m^2-t)^2m^4+4M_Z^2m^4s+4(s+t-m^2)^2M_Z^4\right].
\end{equation}

\begin{figure}
\includegraphics[width=7cm,
  angle=-90]{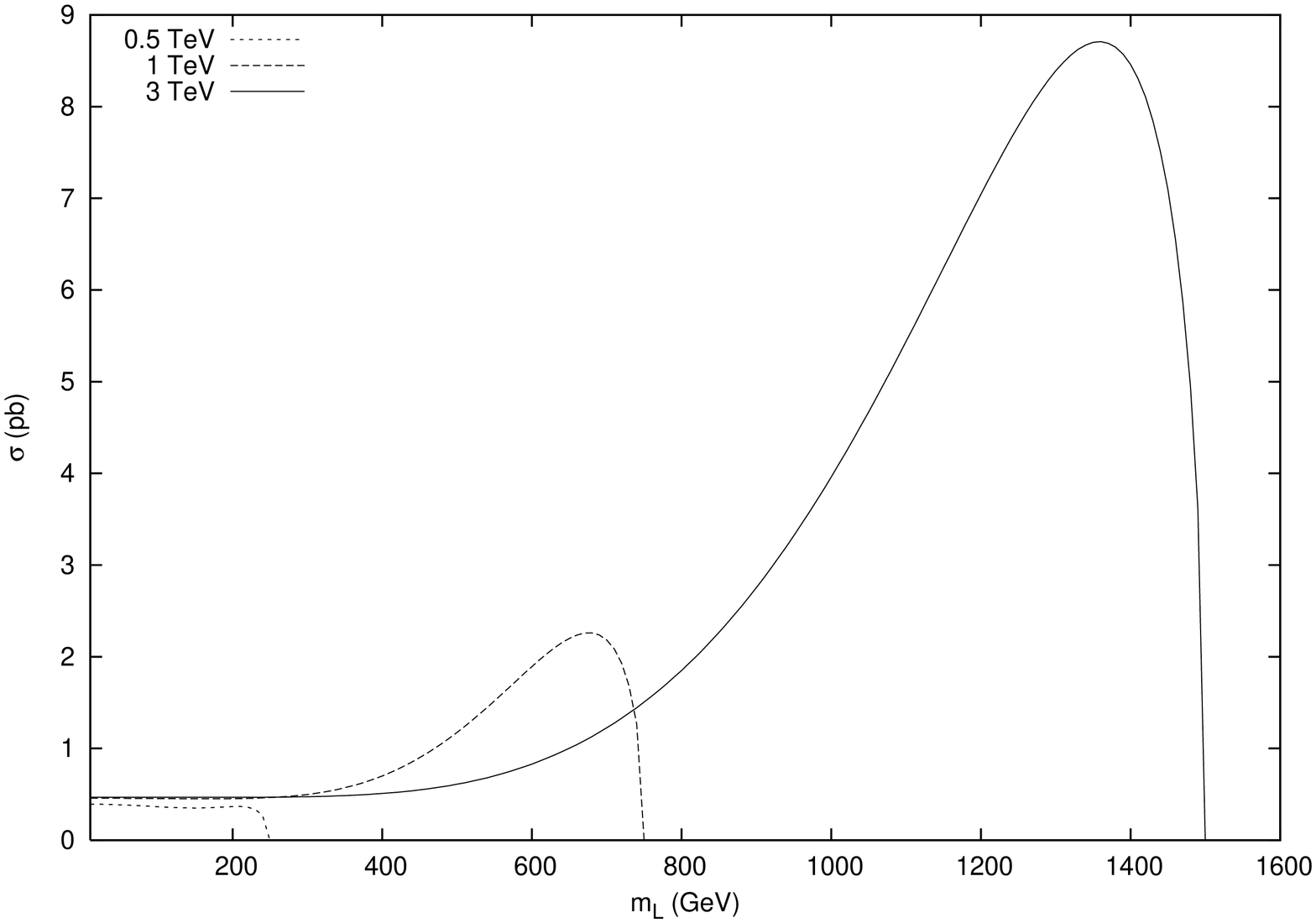}
\caption{\label{fig:f6}The total cross sections as function of the
heavy lepton masses, for the pair production of heavy lepton with
$\sqrt{s}=$0.5 TeV, 1 TeV and 3 TeV.}
\end{figure}

The total cross sections as functions of heavy lepton masses
$m_L$, are displayed in Fig.~\ref{fig:f6}. Signal and background
cross sections depending again on the heavy lepton masses, are
presented in Tables~\ref{tab:t5}, \ref{tab:t6} and \ref{tab:t7} at
0.5, 1 and 3 TeV, respectively. For the pair production of heavy
leptons at linear colliders we expect of order of $10^2-10^3$
signal events for 1250 GeV leptons for the coupling value of
$b_{lLZ}=0.1$. On the other hand, the lower limit of the coupling
which can be probed by pair production at linear colliders is
found to be 0.05.

\begin{figure}
\includegraphics[]{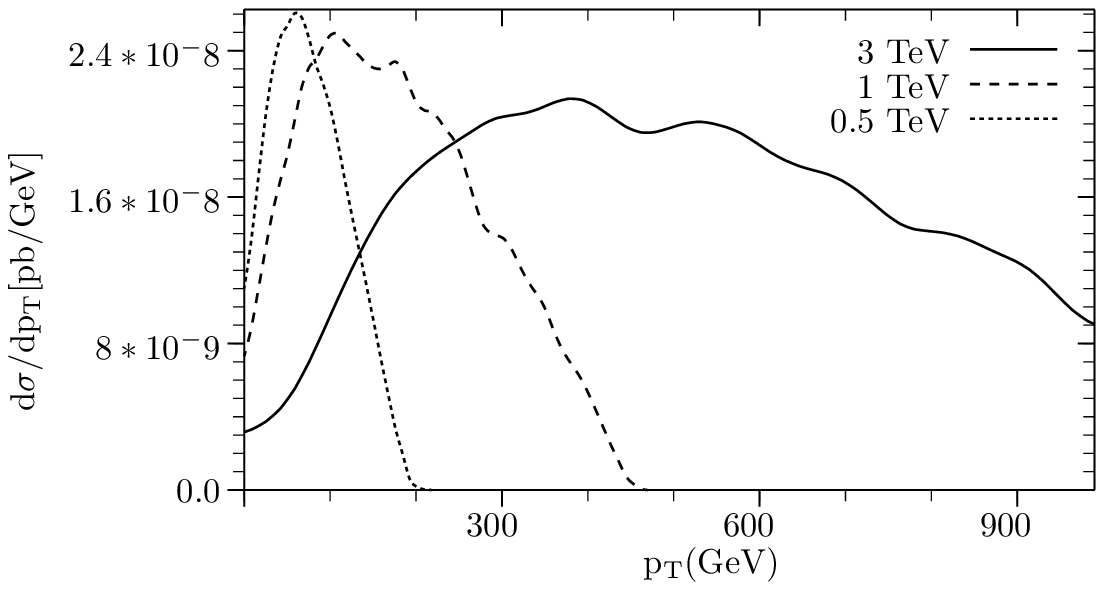}
\caption{\label{fig:f7}$p_T$ distribution of the background at
$\sqrt{s}=$0.5 TeV, 1 TeV and 3 TeV.}
\end{figure}

\begin{figure}
\includegraphics[]{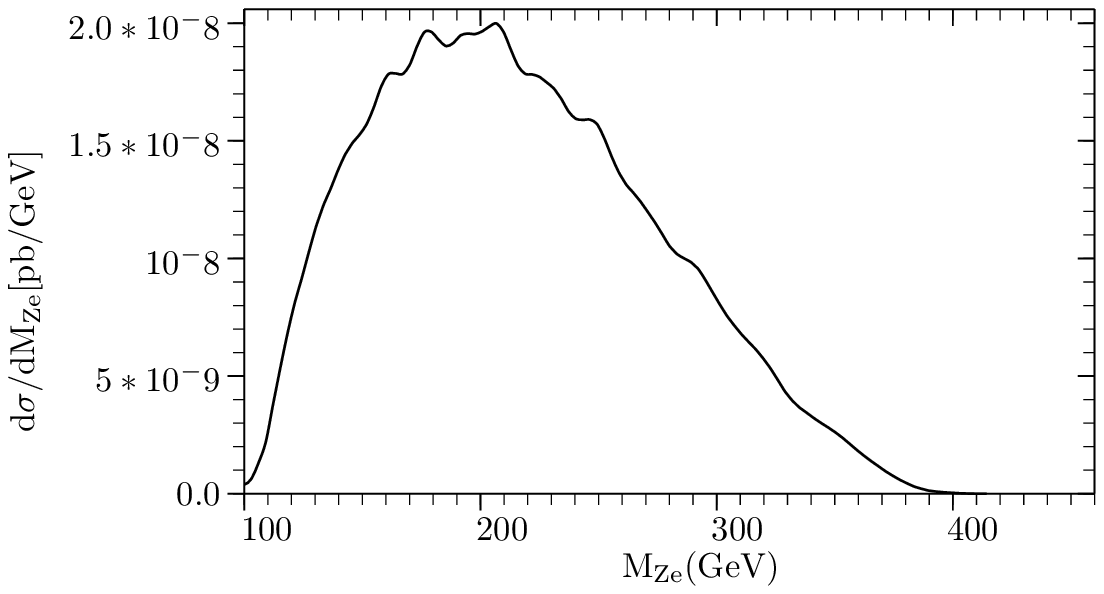}
\caption{\label{fig:f8}The invariant mass distribution of the
$Ze^-$ system for the background at$\sqrt{s}=$0.5 TeV.}
\end{figure}

\begin{figure}
\includegraphics[]{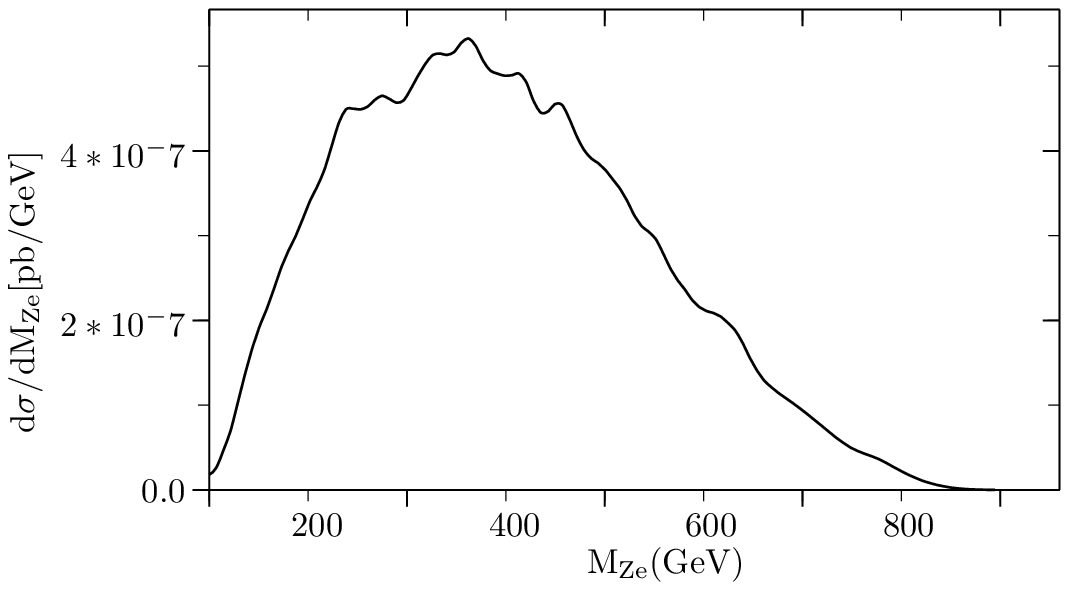}
\caption{\label{fig:f9}The invariant mass distribution of the
$Ze^-$ system for the background at$\sqrt{s}=$1 TeV.}
\end{figure}

\begin{figure}
\includegraphics[]{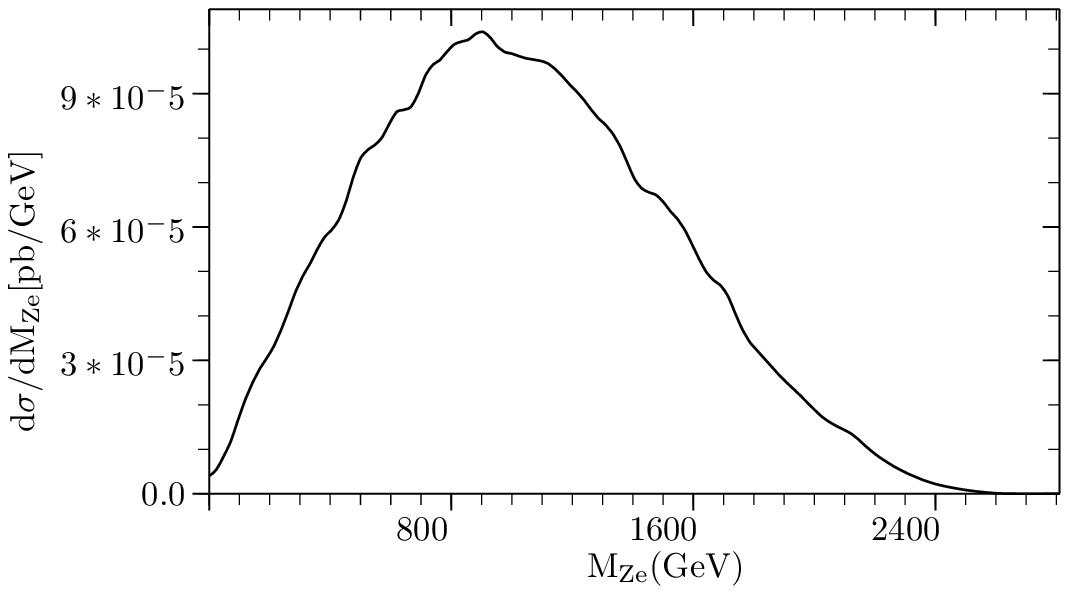}
\caption{\label{fig:f10}The invariant mass distribution of the
$Ze^-$ system for the background at$\sqrt{s}=$3 TeV.}
\end{figure}

We applied an initial cut on the electron and jet transverse
momentum $p_{T}^{e,j}>20$ GeV for the signal and background
analysis. Fig.~\ref{fig:f7} shows the $p_T$ distributions of the
background at the colliders. The distribution of invariant mass
$m_{Ze^-}$ is presented in Figs.~\ref{fig:f8}, \ref{fig:f9} and
\ref{fig:f10} at $\sqrt{s}=$0.5, 1 and 3, respectively.

We have used the high energy package CompHEP for calculations of
background cross sections reported in this study
\cite{Pukhov:1999gg}.

\section{Conclusions}
This study proves that linear colliders can test the existence of
single and pair production of heavy leptons. The production of a
single heavy lepton is more relevant than the pair production.
Namely, in the case of $\sqrt{s}=3$ TeV option, we expect 257
single events for $b_{lLZ}=0.1$ and 26 single events for
$b_{lLZ}=0.032$ for 2750 GeV (which is the upper bound) leptons.
In the case of pair production, for $b_{lLZ}=0.1$ we expect
$10^2-10^3$ events for 1250 GeV (upper value) leptons, while no
pair event can be observed for $b_{lLZ}=0.032$, since the pair
production cross section is suppressed by the fourth power of
mixing couplings.

\begin{acknowledgments} This study was partially supported by Abant
Izzet Baysal University Research fund.
\end{acknowledgments}

\newpage

\begin{table*}[h]
\caption{\label{tab:table1}The main parameters of the future
$e^-e^+$ colliders.  }
\begin{ruledtabular}
\begin{tabular}{ccc}
$e^-e^+$ colliders & $\sqrt{s}$ (TeV) &$\mathcal{L}$($\mathrm{cm}^{-2} \mathrm{s}^{-1}$)\\
\hline
ILC  & 0.5& $10^{34}-10^{35}$\\
CLIC &1.0 & $10^{34}-10^{35}$\\
CLIC &3.0 & $10^{34}-10^{35}$\\
\end{tabular}
\end{ruledtabular}
\end{table*}

\begin{table*}[h]
\caption{\label{tab:t2}The signal and background cross sections
and $S/\sqrt{S+B}$ depending on the heavy lepton masses with
$\sqrt{s}=0.5$ TeV.}
\begin{ruledtabular}
\begin{tabular}{ccccccc}
  $m_L$ (GeV) & $\sigma$ (pb) & $\sigma \times \mathrm{BR}_1$ (pb) & $\sigma \times \mathrm{BR}_1 \times \mathrm{BR}_2$
   (pb) & $\sigma_B \times 10^{-3}$(pb) & $S/\sqrt{S+B}$ & $\Gamma_{\mathrm{Total}}$ (GeV)  \\
  \hline
   100& 1.49 & 0.49 & 0.016 & 0.63 & 39 & 0.008 \\
   200& 1.29 & 0.42 & 0.014 & 1.25 & 36 & 0.589 \\
   300& 0.96 & 0.32 & 0.010 & 1.34 & 30 & 2.159 \\
   400& 0.50 & 0.17 & 0.005 & 1.76 & 20 & 5.183 \\
\end{tabular}
\end{ruledtabular}
\end{table*}

\begin{table*}[h]
\caption{\label{tab:t3}The signal and background cross sections
and $S/\sqrt{S+B}$ depending on the heavy lepton masses with
$\sqrt{s}=1$ TeV.}
\begin{ruledtabular}
\begin{tabular}{ccccccc}
  $m_L$ (GeV) & $\sigma$ (pb) & $\sigma \times \mathrm{BR}_1$ (pb) &
   $\sigma \times \mathrm{BR}_1 \times \mathrm{BR}_2$ (pb) & $\sigma_B \times 10^{-4}$(pb) & $S/\sqrt{S+B}$ & $\Gamma_{\mathrm{Total}}$ (GeV)  \\
  \hline
   100& 1.50 & 0.49 & 0.016 & 0.99 & 40 & 0.008 \\
   300& 1.38 & 0.45 & 0.015 & 2.89 & 38 & 2.159 \\
   500& 1.13 & 0.37 & 0.012 & 2.83 & 35 & 10.146 \\
   700& 0.76 & 0.25 & 0.008 & 3.33 & 28 & 27.834 \\
   900& 0.27 & 0.09 & 0.003 & 5.39 & 16 & 59.108 \\
\end{tabular}
\end{ruledtabular}
\end{table*}

\begin{table*}[h]
\caption{\label{tab:t4}The signal and background cross sections
and $S/\sqrt{S+B}$ depending on the heavy lepton masses with
$\sqrt{s}=3$ TeV.}
\begin{ruledtabular}
\begin{tabular}{ccccccc}
  $m_L$ (GeV) & $\sigma$ (pb) & $\sigma \times \mathrm{BR}_1$ (pb) & $\sigma \times \mathrm{BR}_1 \times \mathrm{BR}_2$ (pb) &
  $\sigma_B \times 10^{-5}$(pb) & $S/\sqrt{S+B}$ & $\Gamma_{\mathrm{Total}}$ (GeV)  \\
  \hline
   250 & 1.49 & 0.49 & 0.016 & 21.06& 40 & 1.22 \\
   750 & 1.40 & 0.46 & 0.015 & 3.97 & 39 & 34.23 \\
   1250& 1.23 & 0.41 & 0.013 & 3.32 & 37 & 158.19 \\
   1750& 0.98 & 0.32 & 0.011 & 4.50 & 33 & 433.67 \\
   2250& 0.65 & 0.22 & 0.007 & 6.35 & 27 & 921.23 \\
   2750& 0.24 & 0.08 & 0.003 & 10.05& 16 & 1681.41 \\
\end{tabular}
\end{ruledtabular}
\end{table*}

\begin{table*}[h]
\caption{\label{tab:t5}The signal and background cross sections
and $S/\sqrt{S+B}$ depending on the heavy lepton masses with
$\sqrt{s}=0.5$ TeV.}
\begin{ruledtabular}
\begin{tabular}{ccccccc}
  $m_L$ (GeV) & $\sigma$ (pb) & $\sigma \times \mathrm{BR}_1$ (pb) & $\sigma \times \mathrm{BR}_1 \times \mathrm{BR}_2$
   (pb) & $\sigma_B \times 10^{-8}$(pb) & $S/\sqrt{S+B}$ & $\Gamma_{\mathrm{Total}}$ (GeV)  \\
  \hline
   100& 0.36 & 0.12 & 0.0039 & 0.005& 20 & 0.01 \\
   150& 0.35 & 0.12 & 0.0038 & 2.49 & 20 & 0.13 \\
   200& 0.37 & 0.12 & 0.0040 & 5.03 & 20 & 0.59 \\
   240& 0.27 & 0.09 & 0.0029 & 1.25 & 17 & 1.22 \\
\end{tabular}
\end{ruledtabular}
\end{table*}

\begin{table*}[h]
\caption{\label{tab:t6}The signal and background cross sections
and $S/\sqrt{S+B}$ depending on the heavy lepton masses with
$\sqrt{s}=1$ TeV.}
\begin{ruledtabular}
\begin{tabular}{ccccccc}
  $m_L$ (GeV) & $\sigma$ (pb) & $\sigma \times \mathrm{BR}_1$ (pb) &
   $\sigma \times \mathrm{BR}_1 \times \mathrm{BR}_2$ (pb) & $\sigma_B \times 10^{-7}$(pb) & $S/\sqrt{S+B}$ & $\Gamma_{\mathrm{Total}}$ (GeV)  \\
  \hline
   100& 0.43 & 0.14 & 0.0047 & 0.004& 22 & 0.01 \\
   200& 0.44 & 0.15 & 0.0048 & 1.51 & 22 & 0.59 \\
   300& 0.56 & 0.18 & 0.0061 & 3.57 & 25 & 2.16 \\
   400& 0.94 & 0.31 & 0.0102 & 5.12 & 32 & 5.18 \\
\end{tabular}
\end{ruledtabular}
\end{table*}

\begin{table*}[h]
\caption{\label{tab:t7}The signal and background cross sections
and $S/\sqrt{S+B}$ depending on the heavy lepton masses with
$\sqrt{s}=3$ TeV.}
\begin{ruledtabular}
\begin{tabular}{ccccccc}
  $m_L$ (GeV) & $\sigma$ (pb) & $\sigma \times \mathrm{BR}_1$ (pb) & $\sigma \times \mathrm{BR}_1 \times \mathrm{BR}_2$ (pb) &
  $\sigma_B \times 10^{-6}$(pb) & $S/\sqrt{S+B}$ & $\Gamma_{\mathrm{Total}}$ (GeV)  \\
  \hline
   250 & 0.47 & 0.16 & 0.0051 & 0.90 & 23 & 1 \\
   500 & 0.61 & 0.20 & 0.0066 & 4.04 & 26 & 10 \\
   750 & 1.51 & 0.50 & 0.0164 & 3.92 & 41 & 34 \\
   1000& 3.96 & 1.31 & 0.0431 & 1.18 & 66 & 81 \\
   1250& 7.79 & 2.57 & 0.0848 & 16.20& 92 & 158 \\
\end{tabular}
\end{ruledtabular}
\end{table*}

\newpage

\end{document}